\documentclass{iopjournal}
\usepackage{graphicx}
\usepackage{amsmath,amssymb}
\usepackage{booktabs}
\usepackage{array}
\usepackage{url}
\usepackage{algorithm}
\usepackage{algpseudocode}

\makeatletter
\renewcommand{\articletype}[1]{{\vspace*{-8mm}\noindent \Large \sf Physica Scripta}

\vspace*{8mm} \noindent\reversemarginpar
\marginpar{\vspace{-3mm} {\color{gray}\hrule} \ \\ Crossmark\\ {\color{gray}\hrule} \ \\ \tiny {\sf RECEIVED} {\small \\ dd Month yyyy}\\ \\ {\sf REVISED} {\small \\ dd Month yyyy}}{\scriptsize \sf{\bfseries \MakeUppercase{#1}}}}
\makeatother

\begin{document}

\articletype{Paper}

\title{Evaluating Quantum Circuit Protection Methods After Compilation: An Empirical Comparison}

\author{Ing Muyleang$^{1}$\,\orcid{0009-0008-0551-1754}, Leanghok Hour$^{2}$\,\orcid{0009-0001-5670-7481} and Youngsun Han$^{1,*}$\,\orcid{0000-0001-7712-2514}}

\affil{$^1$Department of Artificial Intelligence Convergence, Pukyong National University, Busan 48513, Republic of Korea}

\affil{$^2$AI Research Institute, Pukyong National University, Busan 48513, Republic of Korea}

\affil{$^*$Author to whom any correspondence should be addressed.}

\email{youngsun@pknu.ac.kr}

\keywords{quantum cloud computing, quantum circuit obfuscation, circuit cutting, dynamical decoupling, compilation}

\begin{abstract}
Quantum cloud providers can read the circuits they receive and their outputs, yet methods for hiding a circuit are usually evaluated on the circuit as written, before compilation optimises it. We compare three ObfusQate schemes (inverse, composite and delayed gates), ECQCO and Q-MaskCut under one protocol that compiles every protected circuit at optimisation level 3 before measuring it; the goal is to keep the provider from learning the user's algorithm and answer. Q-MaskCut is one example combination of existing techniques: a one-time-pad key, decoy wires, circuit cutting with a random rotation pair across each cut, and Pauli twirling with dynamical-decoupling masking after compilation. We simulate 11 circuits of 3 to 19 qubits without noise and on two noise models, measure the compiled structure and execution time of 23 circuits of 3 to 127 qubits, and test whether the structural disguise survives a second compilation by the provider. Composite and delayed gates lose most of their structural disguise in compilation and leave the output readable; inverse gates keep a high normalised graph edit distance (normGED) only by making execution 7 to 112 times longer; ECQCO's output could not be decoded with the implementation provided by its authors. Q-MaskCut hides the output (mean total variation distance without the key 0.94), and without noise, with exact fragment probabilities, it decodes exactly. Its compiled normGED of 0.88 comes mostly from the masking, which a second compilation removes, leaving 0.42 from the key, decoys and padlocks. With noise, its fidelity is comparable to the original's; it needs 7 to 91 fragment variants per circuit in the simulations (7 to 61 in a run on ibm\_marrakesh) and a classical reconstruction. Protection should therefore be evaluated on the compiled circuit and again after the provider recompiles it.
\end{abstract}

\section{Introduction}

Quantum computing promises to solve some problems faster than classical computing, and current quantum processors are accessed mostly through cloud services~\cite{Preskill2018quantumcomputingin}. A user writes a quantum circuit; the provider compiles it for its device, executes it and returns the measured counts. The provider therefore sees the whole circuit, which may encode a proprietary algorithm, and its output, which may be a confidential answer.

Several lines of work address this problem. Blind and private quantum computation give information-theoretic guarantees, but they require quantum communication or interaction between the user and the provider, which current cloud services do not offer~\cite{childs2005,5438603,fitzsimons2017}. Obfuscation schemes insert gate sequences that act as the identity, so they change how the circuit looks but not its output~\cite{10.1145/3505253.3505260,10.1145/3733817.3762700}. Locking and split-compilation schemes hide the correct configuration of the circuit behind a key or divide the circuit among several compilations~\cite{j6030027,9643478,11050014,11132682}. ECQCO combines a quantum one-time-pad key with structural obfuscation to hide both the structure and the output~\cite{11367741}. Circuit cutting, developed to run circuits larger than a device~\cite{PhysRevLett.125.150504,10.1145/3445814.3446758}, has also been used to obfuscate a circuit by keeping secret how its fragments fit together~\cite{10821443,11250202}.

These methods are usually evaluated on the circuit as written, with different benchmarks and metric definitions, and rarely on hardware. Compilation, however, maps the circuit onto the device's native gates and connectivity and then optimises it, and an optimiser that removes redundant gates can also remove gates inserted for protection. A protection method that appears strong on the written circuit can therefore be much weaker on the circuit that actually runs. It is not known how much of each method's protection remains after compilation, or whether the protected circuit still returns the correct answer on a device.

The goal we study is to keep a cloud provider from learning the user's algorithm and answer from the circuits and counts it receives; defending against a provider that actively attacks or alters the computation is not our goal. In this paper we evaluate representative protection methods under one protocol, in which every protected circuit is compiled for the device at optimisation level 3 before it is measured. Alongside existing schemes, we evaluate Q-MaskCut, one example combination of existing techniques that hides both the output and the structure; we do not present it as a new method. It differs in two ways from earlier cutting-based obfuscation~\cite{10821443}, which relies on the secrecy of how the fragments fit together and on added dummy fragments. First, the circuit carries a one-time-pad key together with decoy wires and padlocks before it is cut, so that the reconstructed output is hidden as well as the structure. Second, every fragment is masked after compilation, so that the masking is not exposed to the user's optimiser; a second compilation by the provider removes it (section~\ref{sec:recompile}).

The contributions of this paper are as follows.
\begin{itemize}
\item We summarise quantum circuit protection methods by what they protect (structure, output or both), their threat models and their evaluation metrics, and we identify the limitations of these metrics.
\item We define an evaluation protocol in which structural disguise is measured after compilation at the highest optimisation level, output disguise is read together with the correctness of the decoded output, and correctness is checked on hardware against the unprotected circuit.
\item As one example that hides both the output and the structure, we evaluate Q-MaskCut, a combination of a one-time-pad key drawn at random, independently of the output, decoy wires with CZ padlocks, CutQC wire cutting with a random rotation pair across each cut, and Pauli twirling with no-pattern dynamical-decoupling masking inserted after compilation. We give its procedures as algorithms and claim novelty neither for the techniques nor for the combination.
\item We compare ObfusQate's inverse, composite and delayed gates, ECQCO and Q-MaskCut under this protocol: on 11 circuits of 3 to 19 qubits without noise and on two noise models, on the compiled structure and execution time of the 23 circuits of 3 to 127 qubits of the hardware run, and on a Bernstein-Vazirani circuit with 200 random inputs. We also run Q-MaskCut and the original on the 156-qubit device ibm\_marrakesh for 23 circuits of 3 to 127 qubits, and we derive recommendations for evaluating protection methods.
\end{itemize}

The remainder of this paper is organised as follows. Section~\ref{sec:background} describes the threat model, reviews protection methods and circuit cutting, and discusses the evaluation metrics. Section~\ref{sec:method} describes Q-MaskCut, the combination we evaluate alongside the existing schemes. Section~\ref{sec:evaluation} describes the experimental setup and evaluates every method. Section~\ref{sec:discussion} discusses the findings and limitations, and section~\ref{sec:conclusion} concludes the paper. Our source code and evaluation artifacts are publicly available at \url{https://github.com/QCL-PKNU/Circuit-Execution-Masking}.

\section{Background and related work}\label{sec:background}

\subsection{Quantum cloud execution and threat model}\label{sec:threat}
A threat model states who the adversary is, what it can observe and do, and what it tries to learn; a claim of protection holds only with respect to it. In cloud execution, the user submits circuits to a provider that runs them on its device. The goal of protection in this study is to keep the provider from learning the user's algorithm and answer, not to defend against malicious attacks. We therefore consider an honest-but-curious provider: it runs the circuits exactly as submitted and returns the true counts, but it may inspect everything it receives (circuits and counts) to learn the user's algorithm or answer. Recompiling a received circuit is routine for a provider, so we also test whether each method's protection survives a second compilation (section~\ref{sec:recompile}). The user is trusted, holds all secret keys and runs classical pre- and post-processing locally. Some schemes assume that the provider also compiles the circuit, so in every study we compile every circuit with the same optimiser before evaluating it, as a provider's compiler would, at the highest level (optimisation level 3). Where a scheme sends parts of a circuit to different providers, we assume that they do not collude. A provider that alters the submitted circuits or the returned counts is outside this model.

A protection method is evaluated on three properties: \emph{structural disguise} (the circuit the provider runs should not reveal the original circuit), \emph{output disguise} (the counts the provider sees should not reveal the answer) and \emph{functional correctness} (after decoding, the user must recover the original output). A fourth, practical property is the \emph{overhead} in circuit depth, execution time, number of executions and classical post-processing.

\subsection{Existing protection methods}
\paragraph{Cryptographic delegation.} Secure assisted and blind quantum computation hide the user's data and, in the blind setting, the computation itself, with information-theoretic guarantees~\cite{childs2005,5438603,fitzsimons2017}. They rely on the quantum one-time pad (QOTP), in which each qubit is encrypted by random Pauli operators~\cite{892142,PhysRevA.67.042317}, and they require quantum communication or interaction between the user and the provider, which current cloud services do not offer. The practical schemes below use their building blocks, in particular the QOTP.

\paragraph{Obfuscation by inserted gates.} Early work inserts dummy gates before compilation and removes them afterwards~\cite{10.1145/3505253.3505260}. ObfusQate~\cite{10.1145/3733817.3762700} provides four circuit-level obfuscations; we use the three that insert identity-equivalent gate sequences: a gate followed by its inverse (\emph{inverse gates}), an auxiliary composite gate followed by its inverse (\emph{composite gates}), and a sequence placed before a group of original gates and cancelled by its inverse after the group (\emph{delayed gates}). These schemes preserve the function of the circuit throughout, so they hide neither the function nor the output from the provider. Das and Ghosh~\cite{11250247} insert a random reversible circuit before compilation and append its inverse afterwards, and OPAQUE~\cite{11014313} hides phase-gate angles behind a key. Both corrupt the function seen by the compiler, but the restored circuit that is executed computes the original output. Theoretical results limit what obfuscation of quantum circuits can achieve in general~\cite{alagic2016quantum}.

\paragraph{Locking and split compilation.} Quantum logic locking~\cite{j6030027} makes the correct function depend on a key. E-LoQ~\cite{11050014} makes selected and dummy gates controlled by a key qubit, so that only the user's key restores the function. Split compilation sends parts of a circuit to several compilers, or to one compiler at different times, so that no single compilation sees all of it~\cite{9643478}. TetrisLock~\cite{11132682} inserts a random reversible circuit and its inverse, and splits the result along an interlocking boundary into two segments that are compiled separately. An oracle-guided attack has recently been shown to recover circuits protected by split compilation~\cite{11310925}.

\paragraph{Encrypted-state compilation.} ECQCO~\cite{11367741} combines output obfuscation (a QOTP key updated gate by gate through the circuit) with structural obfuscation (identity sequences inserted into idle positions, following the concept of dynamical decoupling), so that both the structure and the output are hidden; the user decrypts the output with the final key.

\paragraph{Output hiding by injected flips.} SPYCE~\cite{patel2023toward} hides both the structure and the output by injecting bit flips at the end of the circuit and removing them with a classical decoder, an approach close in spirit to the key of Q-MaskCut.

\subsection{Circuit cutting as a means of protection}
Circuit cutting splits a circuit into smaller fragments that are run separately, and the output is rebuilt classically from the fragment results. Wires~\cite{PhysRevLett.125.150504,10.1145/3445814.3446758,Lowe2023fastquantumcircuit,11131294} or gates~\cite{Mitarai_2021} can be cut, trading quantum for classical resources~\cite{PhysRevX.6.021043}. A wire cut replaces the wire by a sum over measurement and preparation bases, so a circuit with $K$ cuts is rebuilt from a sum of $4^K$ terms over fragment variants. Cutting was designed to run circuits larger than a device, but it also gives each fragment only part of the computation. Typaldos et al.~\cite{10821443,11250202} use this for obfuscation: the user keeps secret how the fragments are joined, and dummy fragments increase the number of circuits a provider would have to consider. Their method hides the structure through the secrecy of the cut, but the fragments preserve the function, so the reconstructed output is the original one; we found no public implementation and compare it qualitatively only. However, fragment metadata can still reveal the family of the algorithm~\cite{punch2026post}, so cutting alone is not complete protection.

\subsection{Evaluation metrics and their limitations}
Structural disguise is measured in ECQCO~\cite{11367741} by the normalised graph edit distance (normGED)~\cite{6313167} between the gate-dependency graphs of the original and the protected circuit,
\begin{equation}
\mathrm{normGED}(A,B)=\frac{\mathrm{GED}(G_A,G_B)}{\max(|V_A|,|V_B|)+\max(|E_A|,|E_B|)}.
\label{eq:normged}
\end{equation}
ECQCO does not specify how a circuit is converted into a graph; we use one node per gate and an edge from each gate to the next gate on each of its qubits. The nodes carry no labels: the edit cost counts only inserted and deleted nodes and edges, so neither the gate type nor the qubit enters the distance (they are used only to choose the starting node matchings of the search). A normGED of 0 means identical structure, and 1 means nothing in common.

OPAQUE, for example, shows that its obfuscated output read without correction does not reveal the dominant solution states, whereas the output corrected with its decoder is close to the original~\cite{11014313}; we use the same two views: the output without the key and the decoded output. Output disguise is measured by the total variation distance (TVD) between the output of the protected circuit and that of the original circuit, in the form used by ECQCO, TetrisLock, E-LoQ and OPAQUE~\cite{11367741,11132682,11050014,11014313},
\begin{equation}
\mathrm{TVD}=\frac{\sum_{i=0}^{2^n-1}\left|y_{\text{Q-MaskCut},i}-y_{\text{origin},i}\right|}{2N},
\label{eq:tvd}
\end{equation}
where $N$ is the number of shots and $n$ the number of output qubits, so that there are $2^n$ possible outcomes. $y_{\text{Q-MaskCut},i}$ and $y_{\text{origin},i}$ are the numbers of shots with outcome $i$ for the protected circuit and for the original circuit, respectively. $y_{\text{Q-MaskCut},i}$ is the obfuscated output, that is, what the provider sees without the key. As in the experiments of TetrisLock~\cite{11132682}, $y_{\text{origin},i}$ is the correct, theoretical output of the original circuit ($y_{\text{origin},i}=Np_i$, with $p$ its exact distribution). A TVD close to 1 means that the output does not reveal the original one, and a TVD close to 0 means that it matches it. Functional correctness is measured by the Hellinger fidelity between the decoded and the ideal output.

\paragraph{Limitations of these metrics.} First, normGED computed on the written circuit ignores compilation, which can cancel inserted gates. Second, normGED rewards any added gates, so a scheme can score highly by making the circuit much longer; it must therefore be read together with depth and execution time. Third, a high TVD does not distinguish an output that is hidden but recoverable from one that is simply wrong, so it must be read together with the correctness of the decoded output. Fourth, neither metric is a formal security guarantee: both measure how different the circuit and its output look, not what a provider could infer with a dedicated analysis.

\section{Q-MaskCut: an example combination of existing techniques}\label{sec:method}

\subsection{Overview and design rationale}
\begin{figure}[!t]
\centering
\includegraphics[width=\textwidth]{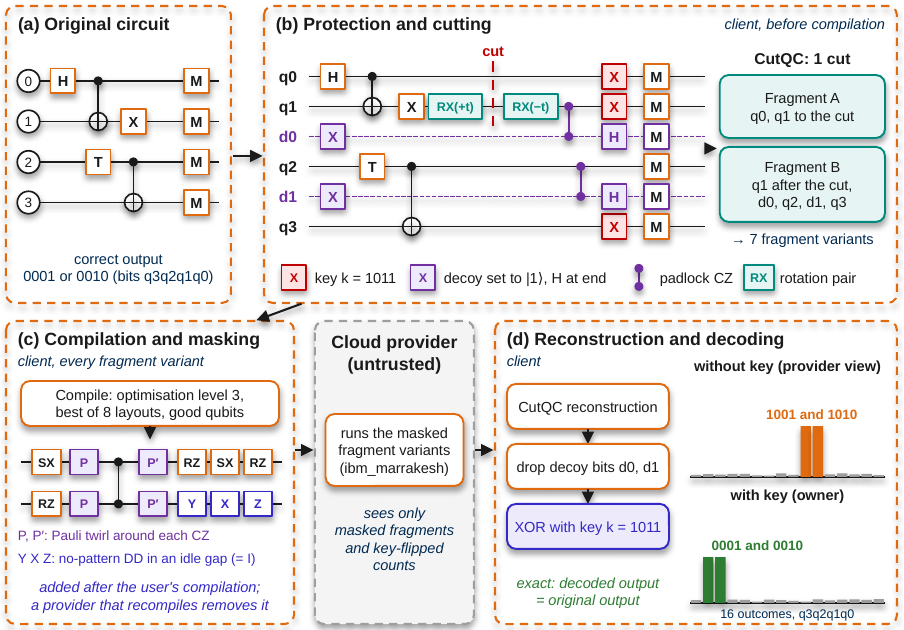}
\caption{Q-MaskCut workflow (illustration). (a) Original four-qubit circuit. (b) Protection and cutting on the client before compilation: key $X$ gates ($k=1011$), two decoy wires set to $|1\rangle$, each linked to a data wire by a CZ padlock and ended with a Hadamard, and one cut with the rotation pair RX$(+t)$/RX$(-t)$, which gives two fragments and seven fragment variants. (c) Every fragment variant is compiled and then masked: Pauli twirl gates ($P$, $P'$) around each two-qubit gate and no-pattern dynamical-decoupling pulses whose product is the identity ($Y$, $X$, $Z$) in idle gaps. The provider runs only the masked fragment variants; a provider that recompiles them removes the masking (section~\ref{sec:recompile}). (d) The client rebuilds the output with CutQC, drops the decoy bits and removes the key by XOR; without the key the peaks are at 1001 and 1010, with the key at the correct 0001 and 0010 (bits $q_3q_2q_1q_0$). The circuit, key and histograms only illustrate the flow; the measured results are in section~\ref{sec:evaluation}.}
\label{fig:walkthrough}
\end{figure}

Figure~\ref{fig:walkthrough} illustrates the flow of Q-MaskCut on a small four-qubit example; the circuit, key and values drawn there illustrate the steps and are not measured results. On the client, the circuit is protected with a key, decoy wires and padlocks (algorithm~\ref{alg:protect}) and cut into fragments with CutQC~\cite{10.1145/3445814.3446758}. A random rotation pair is placed across each cut wire, and every fragment variant is compiled for the device and masked after compilation (algorithm~\ref{alg:mask}). The provider runs only the masked fragment variants. The client then rebuilds the output from the fragment counts, drops the decoys and removes the key (algorithm~\ref{alg:decode}). Before any device time is used, every fragment variant is simulated without noise, and the rebuilt, decoded output must match the ideal output within sampling error.

Each step uses an existing technique and serves one property of section~\ref{sec:threat}. The one-time-pad key~\cite{892142,PhysRevA.67.042317} provides output disguise: the counts the provider receives are bit-flipped by a key it does not hold. The decoy wires and padlocks add qubits and gates that are not part of the computation, so the data wires cannot be identified by counting qubits. A final Hadamard on each decoy makes it read 0 and 1 with equal probability, like a data wire, so the decoys cannot be identified from their marginal output statistics; correlations between output bits can still reveal them (section~\ref{sec:discussion}). Wire cutting~\cite{10.1145/3445814.3446758} gives each fragment only part of the computation, and the rotation pair across each cut changes the content of every fragment without changing the reconstructed output. Pauli twirling~\cite{PhysRevA.94.052325} and no-pattern dynamical decoupling~\cite{PhysRevLett.82.2417,PhysRevLett.94.060502} are inserted after compilation, so they change the structure of the circuit that actually runs.

\subsection{Step 1: one-time-pad key, decoy wires and padlocks}
Algorithm~\ref{alg:protect} adds three layers of protection to the written circuit. Two decoy wires are inserted at random positions among the data wires, so that data and decoy wires are interleaved, and each decoy is prepared in $|1\rangle$. Each decoy is linked to a data wire by a CZ padlock. Because the decoy is in $|1\rangle$, the padlock acts only as a $Z$ on the data wire and does not entangle the two: the decoy stays in a product state, and the padlock adds a two-qubit gate to the structure, not a correlation. The padlock is placed at the earliest point of that wire from which its $Z$, pushed through the rest of the circuit, remains a Pauli operator (a Pauli frame). Any bit flip this frame causes at the measurement is added to the key, so the output remains exactly correctable. A one-time-pad key is then drawn: each data wire is selected with probability $1/2$ (at least one), and on each selected wire an $X$ is placed at the earliest point from which it reaches the measurement as a Pauli that flips at least one output bit and changes the key. The key $k$ is the resulting set of flipped output bits, which the user removes with an XOR after reconstruction. The key is drawn once, at random and independently of the circuit's output, which the user does not know in advance. It is not uniform over all bitstrings: the $X$ on a selected wire can flip several output bits, and the padlock frames add fixed flips, so the distribution of $k$ depends on the circuit. Not every bit flip hides the output equally well: a flip that maps the most likely outcomes onto each other leaves the distribution almost unchanged, so the TVD a random key reaches varies from circuit to circuit (section~\ref{sec:tvd}). Finally, a Hadamard is applied to each decoy at the end of the circuit, so that every decoy reads 0 and 1 with probability $1/2$, as most data wires do; the decoys are marginalised out after reconstruction.

\begin{algorithm}[!t]
\caption{Protect: key, decoys and padlocks}\label{alg:protect}
\begin{algorithmic}[1]
\Require circuit $C$ on $n$ qubits; number of decoys $d=2$; seed $s$
\Ensure protected circuit $C_p$; key $k\in\{0,1\}^n$; decoy wires $D$
\State place the $n$ data wires at random positions among $n+d$ wires; the other $d$ wires form $D$
\State prepare each decoy $q\in D$ with $X$; copy $C$ onto the data wires; $k\gets 0^n$ \Comment{decoy in $|1\rangle$}
\For{each decoy $q\in D$}
  \State choose a data wire $j$; find the earliest point from which $Z_j$, propagated to the end, stays a Pauli
  \State insert padlock $\mathrm{CZ}(q,j)$ there; $k \gets k \oplus$ (output bits flipped by the frame)
\EndFor
\State select each data wire with probability $1/2$ (at least one) \Comment{random key, independent of the output}
\For{each selected data wire $j$}
  \State insert $X_j$ at the earliest point from which it reaches the output as a Pauli that changes $k$
  \State $k \gets k \oplus$ (output bits flipped by this $X$)
\EndFor
\State apply $H$ to each decoy $q\in D$ at the end \Comment{decoys read 0 and 1 equally}
\State \Return $C_p$, $k$, $D$
\end{algorithmic}
\end{algorithm}

\subsection{Step 2: wire cutting, rotation pairs and per-fragment compilation}
The protected circuit is cut with CutQC. Circuits of up to 40 wires are cut into two balanced fragments with at most two cuts (one more cut is allowed when no balanced solution exists); wider circuits are cut into fragments of at most 27 wires. In the hardware run, fragments of circuits below 40 qubits were limited to 20 wires, so bv\_n19, cat\_state\_n22 and ghz\_state\_n23 were cut into three fragments (table~\ref{tab:circuits}). Within these limits, CutQC chooses the cut positions with the fewest cuts. Wires that never interact with the rest of the circuit form separate pieces and run uncut. Each fragment becomes a set of variants in which the cut wires are measured or prepared in the bases required by the reconstruction. A cut wire is measured in three bases and prepared in four states, so a fragment with $a$ measured and $b$ prepared cut wires gives $3^a4^b$ variants. The number of variants of a circuit is the sum over its fragments; for example, one cut gives $3+4=7$ variants.

A random rotation pair is added across each cut wire: $R_X(\theta)$ at the end of the wire in the upstream fragment and $R_X(-\theta)$ at its start in the downstream fragment, with $\theta$ drawn uniformly from $[0,2\pi)$. The pair is the identity on the uncut circuit,
\begin{equation}
 R_X(-\theta)\,R_X(\theta) = I_2 ,
\label{eq:rxpair}
\end{equation}
so the reconstruction remains exact, but the pair changes what each fragment computes and measures. Every variant is compiled by Qiskit~\cite{javadi2024quantum} for the device at optimisation level 3. All variants of a fragment share one layout: the best of eight compilations of that fragment by estimated success probability, on a coupling map from which qubits and links with readout error above 5\%, single-qubit ($\sqrt{X}$) error above 1\% or two-qubit error above 5\% are removed.

\subsection{Step 3: Pauli twirling and no-pattern dynamical decoupling after compilation}
Algorithm~\ref{alg:mask} masks every compiled variant. The variant is scheduled with the device's gate times, so that idle periods become explicit. Around every two-qubit gate, a random Pauli pair is inserted before the gate and the matching correction after it (Pauli twirling~\cite{PhysRevA.94.052325}). The twirl is time-aware: an $X$ is used only where the idle time can hold the pulse, and a virtual $Z$ elsewhere, so the twirl adds no execution time. Every idle gap of the qubits used by the circuit is then filled with up to eight pulse pairs, each an $X$ about a random axis, with no fixed sequence. This is dynamical decoupling~\cite{PhysRevLett.82.2417,PhysRevApplied.20.064027} with randomly chosen pulses~\cite{PhysRevLett.94.060502,PhysRevLett.97.150501}, which we call no-pattern dynamical decoupling. The pulses of each gap multiply to the identity, and this is checked. Because these operations are inserted after compilation, a provider that runs the circuit as submitted does not optimise them away, and because they fill idle periods, they add almost no execution time. A provider that compiles the submitted circuit again, however, removes them (section~\ref{sec:recompile}). Twirling and dynamical decoupling are also known to suppress errors, so they can affect the fidelity as well as the structure.

\begin{algorithm}[!t]
\caption{Mask after compilation: Pauli twirl and no-pattern DD}\label{alg:mask}
\begin{algorithmic}[1]
\Require compiled variant $V$; device target $T$ (gate durations); at most $m=8$ pulse pairs per gap; seed $s$
\Ensure masked variant $V_m$ that computes the same output as $V$
\State $V \gets \textsc{Schedule}(V,T)$ \Comment{idle time becomes explicit delays}
\For{each two-qubit gate $G$ on qubits $(a,b)$}
  \State draw $(P_a,P_b)\in\{I,X,Z\}^2$; use $X$ only where the adjacent idle time holds an $X$ pulse
  \State $(P'_a,P'_b) \gets G\,(P_a\otimes P_b)\,G^\dagger$ \Comment{matching correction}
  \State insert $P_a\otimes P_b$ before $G$ and $P'_a\otimes P'_b$ after $G$
\EndFor
\For{each idle gap of length $\tau$ on a qubit used by the circuit}
  \State $r \gets \min(m,\lfloor \tau/(2t_X)\rfloor)$ \Comment{pairs that fit in the gap}
  \For{$i=1$ to $r$}
    \State draw $\phi$; insert $R_z(\phi)\,X\,R_z(-\phi)$ twice \Comment{random axis, pair = identity}
  \EndFor
\EndFor
\State check that the pulses of every gap multiply to the identity
\State \Return $V_m$
\end{algorithmic}
\end{algorithm}

\subsection{Step 4: reconstruction and decoding on the client}
Algorithm~\ref{alg:decode} rebuilds the output. The fragment counts are converted to probability vectors, sorted by measurement basis and combined by CutQC's reconstruction over the $4^K$ terms. With sampled counts, the reconstruction can give negative quasi-probabilities; as in CutQC, they are replaced by the nearest probability distribution~\cite{PhysRevLett.108.070502}, which is rescaled to sum to 1 before the TVD and the Hellinger fidelity are computed. The decoy wires are then marginalised out and the key is removed by an XOR, which gives the corrected answer. For outputs of up to 29 wires, the whole distribution is rebuilt. For wider circuits, the full distribution does not fit in memory, so the reconstruction computes only the probability of each correct outcome or, for ising\_n34 and ising\_n42 (whose output is spread over many outcomes), the marginal distribution of each block of 12 qubits; knn\_n25 and ising\_n26 are below this width and are rebuilt whole.

\begin{algorithm}[!t]
\caption{Rebuild and decode on the client}\label{alg:decode}
\begin{algorithmic}[1]
\Require counts of every fragment variant; cut solution with $K$ cuts; key $k$; decoy wires $D$
\Ensure corrected output distribution $p$ over the $n$ original qubits
\For{each fragment variant}
  \State convert its counts to a probability vector; sort it by measurement basis
\EndFor
\State $\tilde p \gets$ CutQC reconstruction over the $4^K$ terms \Comment{distribution over all wires}
\State $\tilde p \gets$ marginal of $\tilde p$ over the data wires \Comment{drop the decoys $D$}
\For{each outcome $x$}
  \State $p(x\oplus k) \gets \tilde p(x)$ \Comment{remove the key}
\EndFor
\State \Return $p$
\end{algorithmic}
\end{algorithm}

\subsection{Worked example: bell\_n4 on ibm\_marrakesh}
\begin{figure}[!t]
\centering
\includegraphics[width=\textwidth]{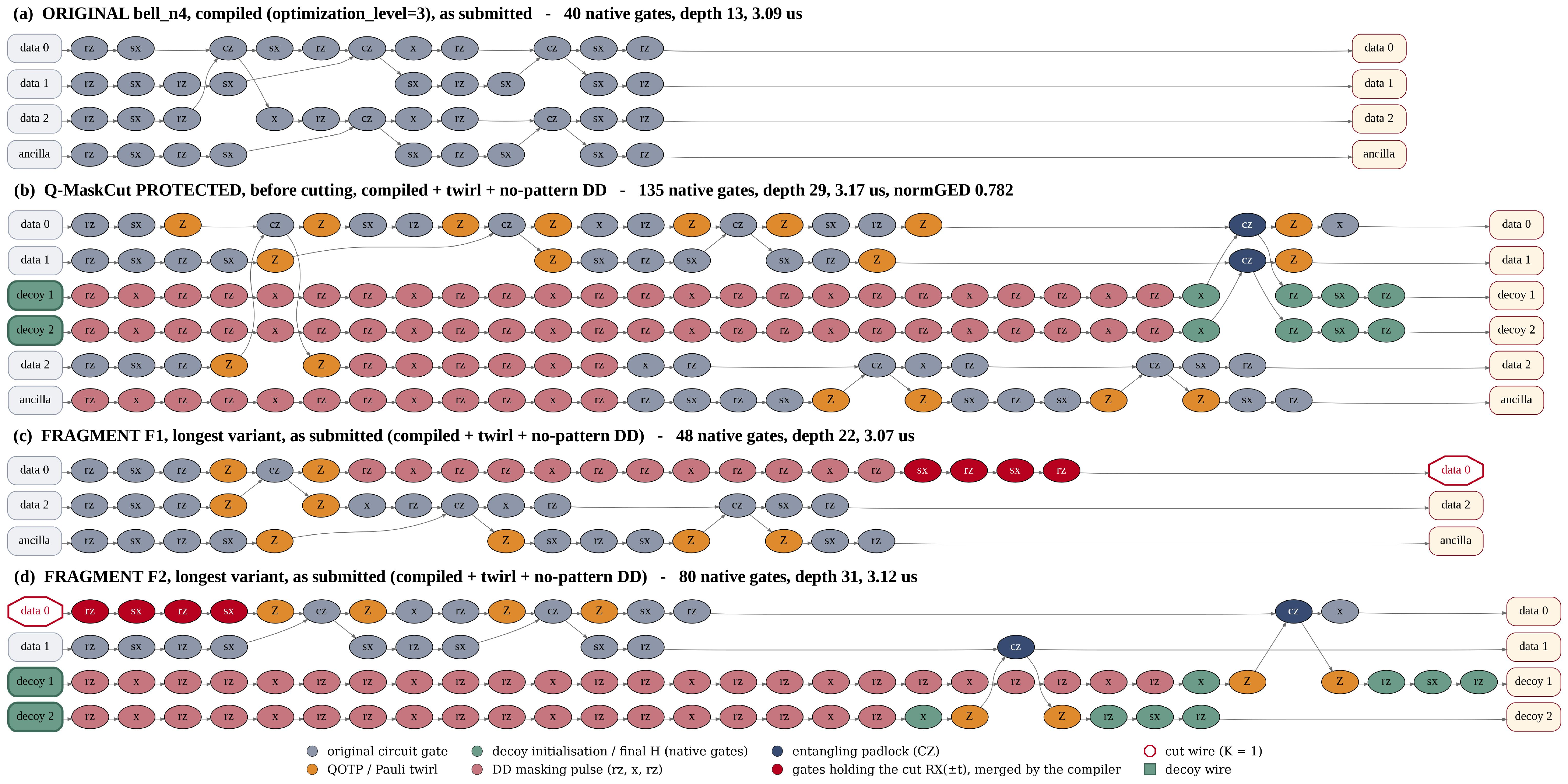}
\caption{Gate-dependency graphs of bell\_n4 on ibm\_marrakesh (full method). (a) Original circuit as compiled; (b) protected circuit before cutting, compiled with the after-compile twirl and masking; (c, d) the longest-running variant of each fragment, as submitted. Octagons mark the cut wire; red nodes are the single-qubit gates that hold the rotation pair across the cut after compilation; pink nodes are the masking pulses, drawn as their native gates $R_z$, $X$, $R_z$.}
\label{fig:dag}
\end{figure}

Figure~\ref{fig:dag} shows the gate-dependency graphs of bell\_n4 at each stage, obtained from the circuits submitted in the hardware run of section~\ref{sec:hardware} (full method) and checked gate by gate against the stored run. The original compiled circuit has 40 native gates (a). The protected circuit before cutting (with the key, the decoys, the padlocks and the decoy Hadamards; this run used the best-TVD key of section~\ref{sec:hardware}), compiled and masked on the same device, has 135 native gates and a normGED of 0.782 with respect to the original (b). The cut leaves one cut wire, data 0, which ends fragment F1 (c) and starts fragment F2 (d); the compiler merges the rotation pair across the cut into the single-qubit gates at the end of F1 and the start of F2. F2 holds both decoys and both padlocks, and the decoys' idle time is filled by the no-pattern pulses. On this short circuit, no idle gap is long enough for an $X$ pulse, so every twirl gate is a virtual $Z$.

\section{Evaluation}\label{sec:evaluation}

\subsection{Experimental setup}\label{sec:setup}
\paragraph{Studies.} The evaluation has three parts. \emph{Tier 1} simulates 11 circuits of 3 to 19 qubits (bv\_n3, bv\_n4, bell\_n4, qpe\_n9, adder\_n10, seca\_n11, multiply\_n13, bv\_n14, qf21\_n15, qec9xz\_n17 and bv\_n19) end to end, without noise and on the FakeMarrakesh and FakeMontrealV2 noise models of Qiskit Aer, with 8192 shots per circuit and per fragment variant. \emph{Tier 2} measures the compiled structure, depth and execution time of the 23 circuits of 3 to 127 qubits of the hardware run (table~\ref{tab:circuits}), compiled for FakeMarrakesh (circuits from QASMBench~\cite{10.1145/3550488}, except bv\_n3, bv\_n4 and the two Deutsch-Jozsa circuits, which are from MQT Bench~\cite{quetschlich2023mqt}; figure~\ref{fig:normged}). These circuits are not simulated, and since their output is not evaluated, Q-MaskCut uses a random key there. The \emph{BV-3 study} runs the three-qubit Bernstein-Vazirani circuit bv\_n3 (secret 11) on 20 iterations of 10 random input states (each input a random $U3$ on every qubit) on FakeMontrealV2, with 1024 shots, optimisation level 3 and transpiler seed 42. Every circuit is scheduled with the device's gate times before it is simulated, so that idle-time noise is charged to every method alike; every method is run 200 times, and we report the mean and standard deviation. In addition, ECQCO's four benchmark algorithms (Bernstein-Vazirani, Grover, QAOA and Shor), in the seven-qubit versions provided by ECQCO's authors, are used only for the comparison with ECQCO's published values (figure~\ref{fig:normged5}). Every circuit is normalised once, in the same way for every method: measurements, barriers and resets are removed, and gates on more than two qubits are decomposed.

\paragraph{Baselines.} ObfusQate's inverse, composite and delayed gates are produced with its released implementation. Its delayed gates did not pass our operator-equivalence check on any of the 5 Tier-1 circuits small enough to check (up to 10 qubits; the 6 larger circuits were not checked) and on 195 of the 200 BV-3 runs. The cause is in the released implementation: none of the nine sequences it inserts as identities is the identity (for example, $YSY$ equals $S^\dagger$ up to a phase, $SZS^\dagger$ equals $Z$, and SWAP--$X$--SWAP moves an $X$ to the neighbouring qubit), and the blocks before and after the circuit, which contain the same sequences in different random orders, cancel only in some cases. We report the measured outputs of the implementation as released. ECQCO is produced with the implementation provided by its authors, and we checked its decoding with the same implementation. Running its own pipeline and simulator (10{,}000 shots) and applying its own final key in both bit orders did not recover the original output (TVD 1.0 on bv\_n14 and seca\_n11), because a bitwise key only relabels outcomes and cannot turn ECQCO's near-uniform or single fixed output back into the original distribution. ECQCO's decoding therefore could not be reproduced with the implementation provided by its authors; we contacted the authors about this and had received no reply at the time of writing. We score its output as it comes out (section~\ref{sec:tvd}), so its obfuscated and deobfuscated outputs coincide. ECQCO produced a circuit for 14 of the 23 Tier-2 circuits and for 10 of the 11 Tier-1 circuits (not for bell\_n4); the others are marked ``not supported'' in the figures.

\paragraph{Compilation and metrics.} In Tier 1 and Tier 2 every circuit, original or protected, is compiled by Qiskit at optimisation level 3 with a fixed seed, as the best of eight compilations by estimated success probability, on the coupling map without poor qubits and links of section~\ref{sec:method}; the BV-3 study also uses level 3, with transpiler seed 42. normGED (equation~(\ref{eq:normged})) is always measured between the compiled original and the compiled protected circuit. For Q-MaskCut, this is the whole protected circuit before cutting (with the key, the decoys, the padlocks and the decoy Hadamards), compiled and then masked by algorithm~\ref{alg:mask}. Since the provider receives only the fragments, this value is a proxy for the structural disguise of what the provider sees. The masking that Q-MaskCut adds after compilation is therefore included; a normGED computed on the logical circuit would miss it: for BV-3, Q-MaskCut's normGED is $0.44$ on the logical circuit and $0.896$ on the compiled circuit, whereas composite gates fall from $0.56$ on the logical circuit to $0.00$ after compilation. The GED is computed as a reproducible upper bound (three starting matchings, each refined by a seeded local search and its edit cost counted exactly), which equals the exact GED on small circuits, where the exact value can be computed.

Output disguise is measured by equation~(\ref{eq:tvd}) between the output the provider sees, read without the key, and the correct output ($\mathrm{TVD}_{\mathrm{obf}}$); for Q-MaskCut, this is the output reconstructed from all fragments without removing the key. In the BV-3 study, we report the smaller value $\mathrm{TVD}_{\mathrm{obf,min}}$: the minimum over the individual fragments of the TVD between the data wires that a fragment measures and the corresponding marginal of the correct output, which is what a provider holding a single fragment sees. Correctness is measured by the TVD between the decoded output and the correct output ($\mathrm{TVD}_{\mathrm{deobf}}$), by the Hellinger fidelity of the decoded output and, as an exactness check, by $\mathrm{TVD}_{\mathrm{ideal}}$, the TVD of the output decoded from the exact noiseless fragment probabilities. Execution time is the scheduled duration of the compiled circuit; for Q-MaskCut, it is that of the longest-running fragment. A noisy simulation was skipped (``too large to simulate'') when the number of gates times $2^{\text{qubits}}$ per shot exceeded $3\times10^{8}$; this affects only inverse gates, on the deepest Tier-1 circuits.

\subsection{Structural disguise after compilation}\label{sec:structure}
\begin{figure}[!t]
\centering
\includegraphics[width=\textwidth]{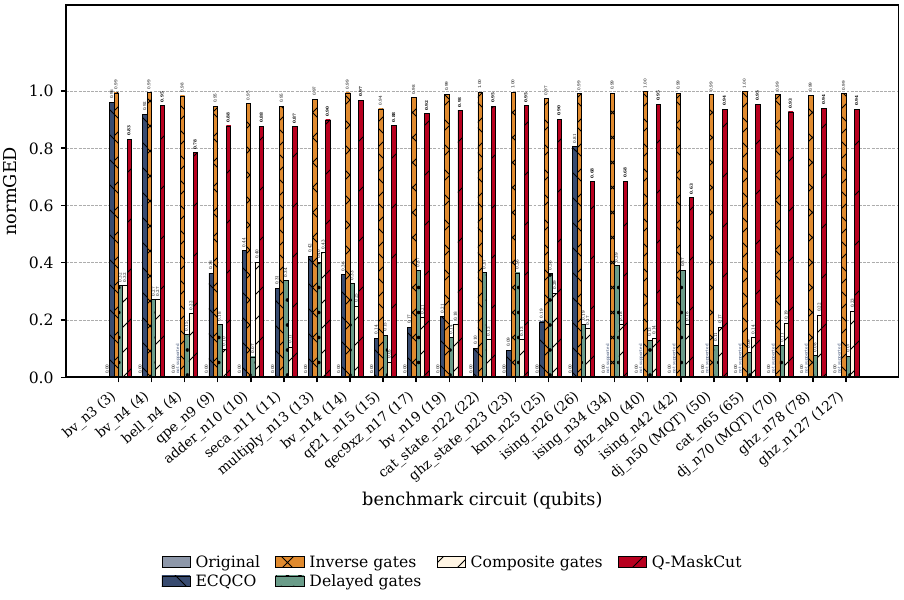}
\caption{normGED after compilation (Tier 2: the 23 circuits of the hardware run). Every circuit is compiled at optimisation level 3 for FakeMarrakesh (best of eight compilations on good qubits); Q-MaskCut includes the masking applied after compilation. ``not supported'': the method's own code gives no circuit.}
\label{fig:normged}
\end{figure}

Figure~\ref{fig:normged} shows normGED after compilation for the 23 Tier-2 circuits. Composite and delayed gates fall to 0.05--0.43 and 0.07--0.40 (means 0.21 and 0.23): the compiler cancels most of the identity sequences they insert, and their compiled circuits are close to the original. ECQCO reaches 0.09--0.96 (mean 0.39) on the 14 circuits for which its implementation produced a circuit. Inverse gates retain 0.94--1.00, but only by making execution 7 to 112 times longer (section~\ref{sec:overhead}). Q-MaskCut reaches 0.63--0.97 (mean 0.88), and 0.83--0.97 on 19 of the 23 circuits, up to 127 qubits; the exceptions are bell\_n4 (0.78) and the Ising circuits of 26 to 42 qubits (0.63--0.68), which are shallow and leave little idle time to mask. On the Tier-1 circuits compiled for FakeMarrakesh, the mean normGED is 0.89 for Q-MaskCut, 0.97 for inverse gates, 0.41 for ECQCO, and 0.23 and 0.25 for composite and delayed gates, respectively.

The ranking follows from where each method inserts its gates. Composite and delayed gates insert identity sequences that the level-3 optimiser recognises and removes. Inverse gates survive because each inserted pair is interleaved with the circuit, but their high normGED comes from added length, which equation~(\ref{eq:normged}) rewards, rather than from hidden structure. Q-MaskCut obtains most of its normGED from pulses inserted into idle periods after compilation, so its value depends on how much idle time the compiled circuit has, and it holds only as long as the provider does not compile the circuit again (section~\ref{sec:recompile}). The BV-3 study shows the same effect on one circuit (figure~\ref{fig:bv3}): after compilation, normGED is $0.00$ for composite gates, $0.23$ for delayed gates, $0.53$ for ECQCO, $0.90$ for Q-MaskCut ($0.45$ without the masking, table~\ref{tab:recompile}) and $0.98$ for inverse gates.

\begin{figure}[!t]
\centering
\includegraphics[width=\textwidth]{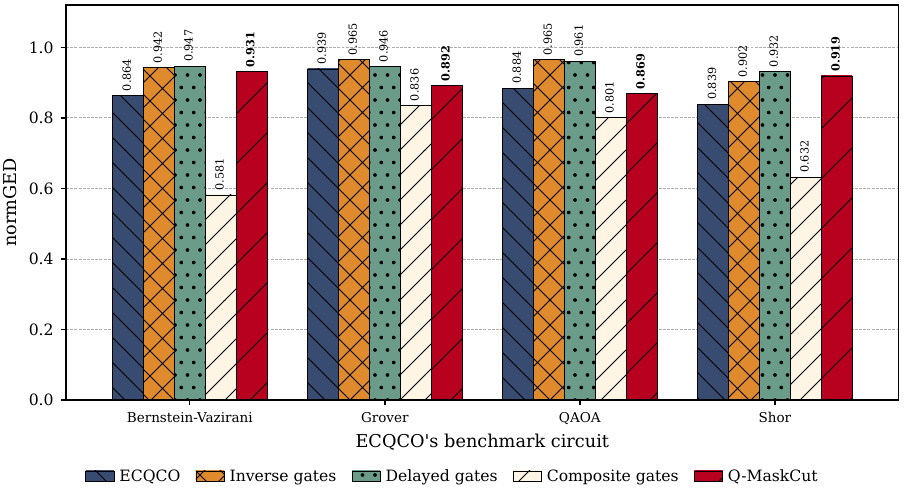}
\caption{normGED on ECQCO's four benchmark circuits. Baselines from~\cite{11367741}; Q-MaskCut (full method) measured after compilation for FakeMarrakesh, masking included.}
\label{fig:normged5}
\end{figure}

On ECQCO's own four benchmark circuits (figure~\ref{fig:normged5}), we compare Q-MaskCut with the normGED values reported in~\cite{11367741} for ECQCO and for inverse, composite and delayed gates. Q-MaskCut is measured with the full method and the same code as Tier 2 (FakeMarrakesh, optimisation level 3, masking included). It reaches 0.87--0.93, within the range of ECQCO (0.84--0.94) and above composite gates (0.58--0.84). It is below inverse and delayed gates (0.90--0.97) on three of the four circuits; on Shor, it is above inverse gates (0.92 against 0.90) and below delayed gates (0.93). These values are those of the circuits as published. When ECQCO's distributed obfuscated files are compiled in the same way as every other circuit in this study, composite gates fall to 0.00--0.48 and delayed gates to 0.09--0.44, consistent with figure~\ref{fig:normged}.

\subsection{Second compilation by the provider}\label{sec:recompile}
The comparison above shows that composite and delayed gates fail because the compiler removes what they insert. The same test must be applied to Q-MaskCut, whose normGED comes mostly from the masking added after compilation. A provider routinely recompiles the circuits it receives, so we removed the delays from every submitted circuit, masking included, and transpiled it again with Qiskit at optimisation level 3 for the same target, then measured normGED against the compiled original with the same code (table~\ref{tab:recompile}).

The masking does not survive a second compilation. On the 23 Tier-2 circuits, the mean normGED is 0.38 before masking, 0.88 after masking and 0.42 after recompilation; on BV-3 it is $0.45\pm0.04$, $0.90\pm0.00$ and $0.48\pm0.02$. On every one of the 53 submitted fragments, the recompiled fragment has at most as many gates as before masking plus 2.4\% of the masking gates, and its normGED with respect to the unmasked fragment falls from 0.81 to 0.32 on average. The optimiser merges the pulse pairs and the twirl Paulis into the neighbouring single-qubit gates, as it does with the identity sequences of composite and delayed gates. Keeping the layout fixed gives the same result (mean 0.42), and recompiling the compiled original itself gives 0.05 with the layout fixed and 0.23 when the compiler may choose a new layout, so the remaining value of 0.42 comes from the key, the decoys and the padlocks, which the optimiser does not remove. Surviving the optimiser is not the same as hiding the algorithm, however: part of this value comes from the decoy wires and the gates they add, which normGED without node labels counts, and the decoys can be identified from correlations between output bits (section~\ref{sec:discussion}).

We applied the same second compilation to every other method (table~\ref{tab:recompileall}). These methods insert their gates before the first compilation, so whatever an optimiser can remove is already gone, and a second compilation changes little: with the layout fixed, inverse gates stay at 0.98, ECQCO at 0.40 and composite and delayed gates at 0.23 and 0.26, against 0.05 for the original. Q-MaskCut is the only method whose normGED falls, because it is the only one that adds gates after compilation. Against a provider that recompiles, Q-MaskCut's structural disguise is therefore that of its logical changes (0.42): above composite and delayed gates, comparable to ECQCO (0.44 against 0.40 on the 14 circuits for which ECQCO produced a circuit) and far below inverse gates. Its output disguise is unaffected, since recompilation preserves the function and the key still flips the output.

\begin{table}
\caption{normGED of Q-MaskCut against the compiled original (Tier 2, FakeMarrakesh): the compiled protected circuit before masking, after the masking of algorithm~\ref{alg:mask} (the value of figure~\ref{fig:normged}), and after the provider removes the delays and recompiles it at optimisation level 3 for the same target.}
\label{tab:recompile}
\centering
\small
\begin{tabular}{lrrr|lrrr}
\toprule
circuit & before & masked & recompiled & circuit & before & masked & recompiled \\
\midrule
bv\_n3 & 0.51 & 0.83 & 0.54 & ghz\_state\_n23 & 0.32 & 0.95 & 0.33 \\
bv\_n4 & 0.68 & 0.95 & 0.59 & knn\_n25 & 0.42 & 0.90 & 0.41 \\
bell\_n4 & 0.46 & 0.78 & 0.50 & ising\_n26 & 0.26 & 0.68 & 0.39 \\
qpe\_n9 & 0.42 & 0.88 & 0.41 & ising\_n34 & 0.33 & 0.68 & 0.32 \\
adder\_n10 & 0.26 & 0.88 & 0.48 & ghz\_n40 & 0.39 & 0.95 & 0.28 \\
seca\_n11 & 0.41 & 0.87 & 0.43 & ising\_n42 & 0.31 & 0.63 & 0.34 \\
multiply\_n13 & 0.36 & 0.90 & 0.40 & dj\_n50 & 0.39 & 0.94 & 0.37 \\
bv\_n14 & 0.73 & 0.97 & 0.65 & cat\_n65 & 0.29 & 0.95 & 0.34 \\
qf21\_n15 & 0.29 & 0.88 & 0.39 & dj\_n70 & 0.35 & 0.93 & 0.37 \\
qec9xz\_n17 & 0.41 & 0.92 & 0.43 & ghz\_n78 & 0.37 & 0.94 & 0.38 \\
bv\_n19 & 0.33 & 0.93 & 0.45 & ghz\_n127 & 0.29 & 0.94 & 0.40 \\
cat\_state\_n22 & 0.27 & 0.95 & 0.37 &  & & &  \\
\midrule
mean & 0.38 & 0.88 & 0.42 & BV-3 (200 runs) & 0.45 & 0.90 & 0.48 \\
\bottomrule
\end{tabular}
\end{table}

\begin{table}
\caption{Mean normGED against the compiled original on the 23 Tier-2 circuits (FakeMarrakesh) after the user's compilation and after the provider removes the delays and compiles the circuit again at optimisation level 3, with a new layout or with the layout fixed. ECQCO: the 14 circuits for which its implementation produced a circuit.}
\label{tab:recompileall}
\centering
\small
\begin{tabular}{lrrr}
\toprule
method & compiled once & recompiled, new layout & recompiled, fixed layout \\
\midrule
original & 0.00 & 0.23 & 0.05 \\
inverse gates & 0.98 & 0.98 & 0.98 \\
composite gates & 0.21 & 0.32 & 0.23 \\
delayed gates & 0.23 & 0.32 & 0.26 \\
ECQCO & 0.39 & 0.47 & 0.40 \\
Q-MaskCut & 0.88 & 0.42 & 0.42 \\
\bottomrule
\end{tabular}
\end{table}

\subsection{Output disguise}\label{sec:tvd}
Following TetrisLock~\cite{11132682}, we measure output disguise with equation~(\ref{eq:tvd}) between the output of the protected circuit as the provider sees it (read without the key) and the correct output of the original circuit; a TVD close to 1 means that the provider's output does not reveal the answer. A high TVD must be read together with the correctness of the decoded output (section~\ref{sec:fidelity}), since an output that is simply wrong also has a high TVD.

Two limitations of this measure should be kept in mind. First, TVD compares output distributions only: it shows whether the counts the provider receives resemble the correct output, not that the circuit is secure against every analysis. Second, Q-MaskCut's key flips output bits, so it changes which bitstrings are produced but not how the output is spread, and the TVD it can reach is bounded by the circuit's own output; for bell\_n4, no flip pattern gives more than about 0.71, and the drawn key gives 0.36 without noise. The TVD values reported by other works are not directly comparable with ours, since they differ in the reference output, the benchmark circuits, the noise model and the number of shots; we therefore do not rank Q-MaskCut against published values.

\begin{figure}[!t]
\centering
\includegraphics[width=\textwidth]{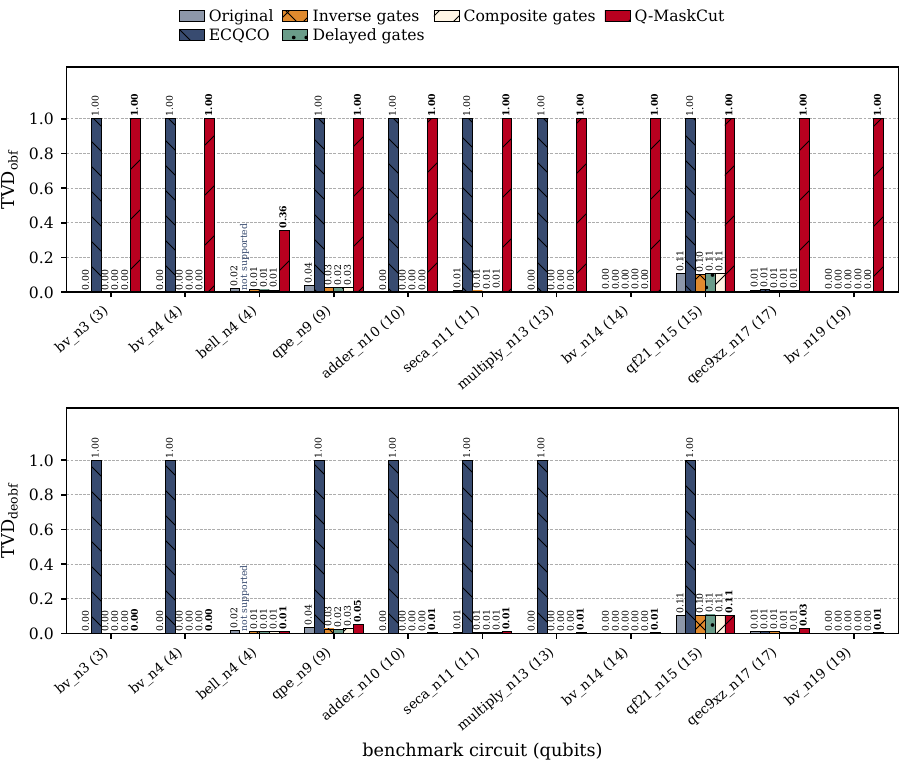}
\caption{Output disguise and recovery, Tier 1, noiseless simulation, 8192 shots. Top: $\mathrm{TVD}_{\mathrm{obf}}$, the TVD between the output the provider sees (no key) and the correct output; higher means more hidden. Bottom: $\mathrm{TVD}_{\mathrm{deobf}}$, the TVD between the output decoded with the key and the correct output; lower means correctly recovered. A method hides the answer usefully only if it is high in the top panel and low in the bottom panel.}
\label{fig:tvd}
\end{figure}

\begin{figure}[!t]
\centering
\includegraphics[width=\textwidth]{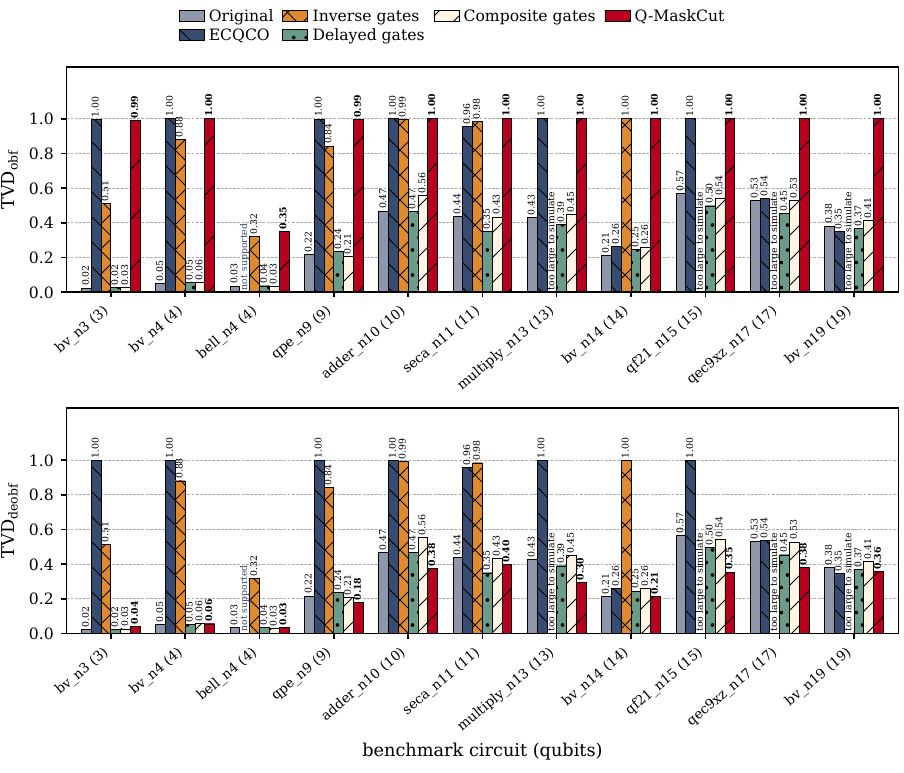}
\caption{Output disguise and recovery with noise: Tier 1 on the FakeMarrakesh noise model, every circuit scheduled so that idle-time noise is charged alike, 8192 shots. Top: $\mathrm{TVD}_{\mathrm{obf}}$; bottom: $\mathrm{TVD}_{\mathrm{deobf}}$. With noise, the reference for the bottom panel is the original's own value, not zero. ``too large to simulate'': above the simulation-cost limit.}
\label{fig:tvdnoise}
\end{figure}

Figure~\ref{fig:tvd} reports $\mathrm{TVD}_{\mathrm{obf}}$ for Tier 1. Without noise, Q-MaskCut's output read without the key has a TVD of 1.00 on 10 of the 11 circuits and 0.36 on bell\_n4 (mean 0.94), while the output decoded from the exact fragment probabilities equals the correct output ($\mathrm{TVD}_{\mathrm{ideal}}\le3\times10^{-15}$ on every circuit). With noise, the mean is 0.94 on both FakeMarrakesh and FakeMontrealV2 (0.35 on bell\_n4). The key is drawn without looking at the output: on 10 of the 11 circuits the drawn key happens to be one of the flips with the largest TVD, whereas on bell\_n4 it reaches 0.36 of the possible 0.71, which shows how much a random key's disguise depends on the circuit. Since the key only flips output bits, the output the provider would see under any other key follows from the same decoded output, so we also average $\mathrm{TVD}_{\mathrm{obf}}$ over keys. This is an unweighted mean over all $2^n-1$ non-zero flip patterns of the output bits (20{,}000 sampled ones for 17 and 19 wires), not weighted by the probability with which algorithm~\ref{alg:protect} draws each key. The mean over keys is 0.94 without noise and 0.94 and 0.93 on FakeMarrakesh and FakeMontrealV2, close to the drawn keys. It is at least 0.99 on 9 of the 11 circuits, 0.95 on qpe\_n9 and $0.38\pm0.15$ on bell\_n4 without noise; single keys that flip only wires with a spread-out output give much less, down to 0.01 on seca\_n11. A single fragment can reveal more than the whole: for bell\_n4, the data wires of one fragment differ from the correct marginal by only 0.25 without noise. The three ObfusQate schemes preserve the circuit's function, so without noise their output equals the original output (mean TVD 0.015, the shot-noise level of the original) and the provider reads the answer. With noise, their TVD equals that of the unprotected circuit (means 0.28--0.32 against 0.30 on FakeMarrakesh, figure~\ref{fig:tvdnoise}), which reflects device noise rather than hiding. The higher values of inverse gates on the noise models (means 0.79 and 0.91) are caused by noise in their much longer circuits, not by hiding. With noise, no method decodes to a TVD of zero. Q-MaskCut's decoded output is at or below the original's on 8 of the 11 circuits on FakeMarrakesh (unrounded values; bell\_n4 is above by 0.0001; mean $\mathrm{TVD}_{\mathrm{deobf}}$ 0.24 against 0.30) and on 6 of 11 on FakeMontrealV2 (0.50 against 0.59), and above it by at most 0.02 on the others. ECQCO's output differs from the correct output (mean 0.70 without noise), but it still differs after decoding with its own final key ($\mathrm{TVD}_{\mathrm{deobf}}=1.00$ on seven circuits, figure~\ref{fig:tvd}, bottom), so it hides the answer from the user as well as from the provider. Q-MaskCut's decoded output is within 0.02 of the unprotected circuit's own 8192-shot output on every circuit (at most 0.11, on qf21\_n15, where the original also has 0.11).

\begin{figure}[!t]
\centering
\includegraphics[width=\textwidth]{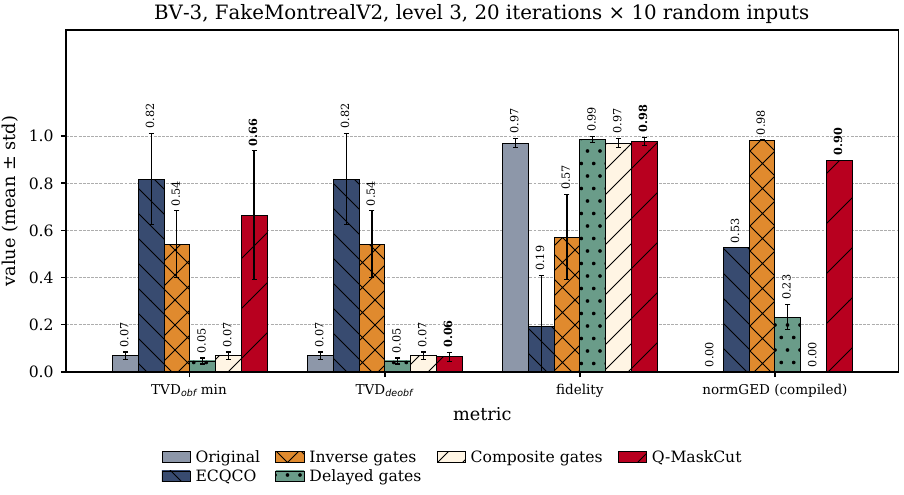}
\caption{BV-3 study: bv\_n3 on FakeMontrealV2, 1024 shots, optimisation level 3, every circuit scheduled so that idle-time noise is charged alike, 20 iterations $\times$ 10 random $U3$ inputs (mean $\pm$ standard deviation over 200 runs). $\mathrm{TVD}_{\mathrm{obf,min}}$: output seen without the key (Q-MaskCut: minimum over fragments); $\mathrm{TVD}_{\mathrm{deobf}}$: decoded output; fidelity of the decoded output; normGED after compilation, including Q-MaskCut's masking.}
\label{fig:bv3}
\end{figure}

\begin{figure}[!t]
\centering
\includegraphics[width=\textwidth]{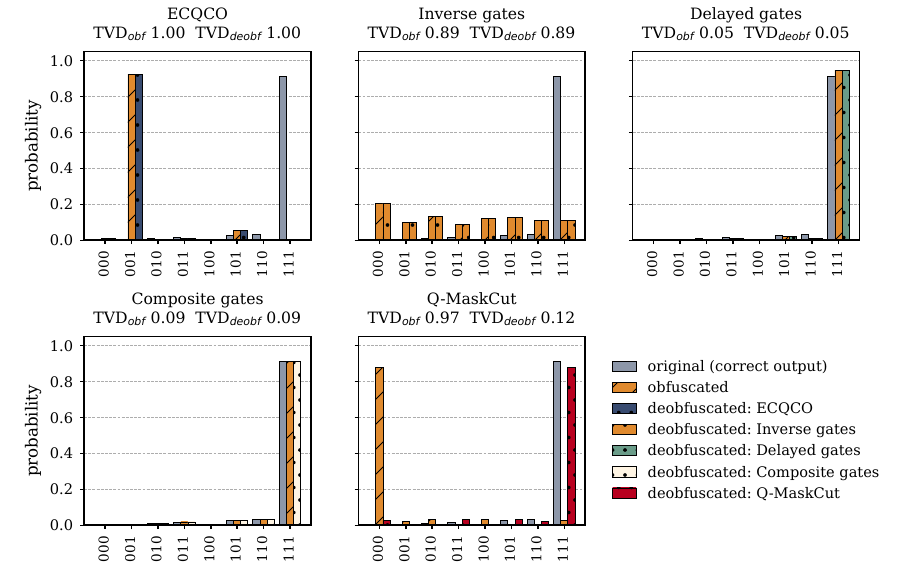}
\caption{BV-3 (secret 11), plain input, FakeMontrealV2, all eight outcomes written $q_2q_1q_0$ ($q_0$ is the flag qubit; the correct output is 111): the original output, the obfuscated output (what the provider sees, without the key) and the deobfuscated output (with the key), for each method; $\mathrm{TVD}_{\mathrm{obf}}$ and $\mathrm{TVD}_{\mathrm{deobf}}$ in each panel title.}
\label{fig:bv3bits}
\end{figure}

On BV-3 with random inputs (figure~\ref{fig:bv3}), with every circuit scheduled so that idle-time noise is charged alike, Q-MaskCut has a per-fragment $\mathrm{TVD}_{\mathrm{obf,min}}$ of $0.66\pm0.27$, against $0.07\pm0.02$ for the original, and it decodes to the original's level ($\mathrm{TVD}_{\mathrm{deobf}}$ $0.06$ against $0.07$; fidelity of the decoded output $0.977\pm0.018$ against $0.969\pm0.020$). Composite gates ($0.07$, fidelity $0.970$) and delayed gates ($0.05$, fidelity $0.985$) leave the output visible. Delayed gates have a slightly higher fidelity than Q-MaskCut, but they hide nothing; the fidelities are thus comparable, and the difference between the methods lies in what the provider sees. Inverse gates reach $0.54$, but their decoded output has the same TVD and a fidelity of only $0.571$, so the output is degraded for the owner too. ECQCO reaches $0.82$, and its decoded output has the same TVD and a fidelity of $0.193$. Figure~\ref{fig:bv3bits} shows all eight outcomes for the plain input (the correct output is 111, including the flag qubit). The obfuscated output of Q-MaskCut is mostly 000 ($\mathrm{TVD}_{\mathrm{obf}}=0.97$), and its deobfuscated output is the correct 111 (probability $0.88$, $\mathrm{TVD}_{\mathrm{deobf}}=0.12$, the noise level of the original). Inverse gates give a near-uniform output, ECQCO gives mostly 001 before and after deobfuscation, and composite and delayed gates show the correct 111 without any key.

\subsection{Functional correctness on the noise models}\label{sec:fidelity}
Without noise, every decodable method returns the correct output: the mean fidelity of the decoded output is 0.996 for the original and for the ObfusQate schemes, and 0.989 (0.955--1.000) for Q-MaskCut, whose reconstruction from 7 to 91 separately sampled fragment variants per Tier-1 circuit adds sampling error. With exact fragment probabilities, its decoded output is exact (section~\ref{sec:tvd}). ECQCO's decoded output has a mean fidelity of 0.30 without noise.

\begin{figure}[!t]
\centering
\includegraphics[width=\textwidth]{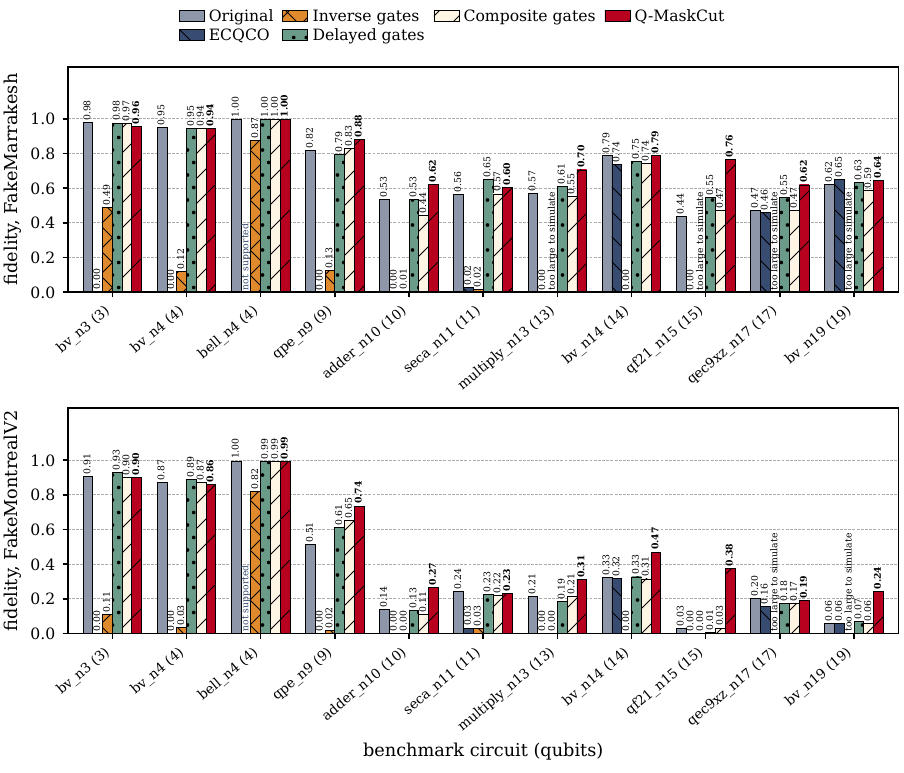}
\caption{Fidelity of the decoded output, Tier 1, on the FakeMarrakesh (top) and FakeMontrealV2 (bottom) noise models, every circuit scheduled so that idle-time noise is charged alike, 8192 shots. ``too large to simulate'': inverse gates above the simulation-cost limit; ``not supported'': ECQCO's code gives no circuit.}
\label{fig:fidelity}
\end{figure}

With noise (figure~\ref{fig:fidelity}), the fidelity of Q-MaskCut is comparable to the original's. Every circuit is scheduled before simulation, so that idle-time noise is charged to every method alike. On FakeMarrakesh, Q-MaskCut is higher by more than 0.02 on 7 of the 11 circuits (the smallest of these margins is 0.025, on bv\_n19) and within 0.02 on the other four (mean 0.775 against 0.702), for example 0.764 against 0.437 on qf21\_n15. On FakeMontrealV2, it is higher by more than 0.02 on 6 circuits and within 0.02 on the other five (mean 0.507 against 0.409), for example 0.736 against 0.511 on qpe\_n9 and 0.377 against 0.030 on qf21\_n15. It is slightly lower on the shortest circuits, such as bv\_n3 (0.958 against 0.976 on FakeMarrakesh), where cutting saves little and the reconstruction adds sampling error. The larger values on the deeper circuits come from shorter executed fragments together with the twirl and pulses, whose contributions we did not separate. Composite and delayed gates stay close to the original, since compilation removes most of what they add, whereas inverse gates fall far below it. Fidelity is therefore not what distinguishes Q-MaskCut: it hides the output (section~\ref{sec:tvd}) while the owner recovers the correct output, and it hides the compiled structure as long as the provider does not recompile it (section~\ref{sec:recompile}).

\subsection{Overhead: time, depth, executions and reconstruction}\label{sec:overhead}
\begin{figure}[!t]
\centering
\includegraphics[width=\textwidth]{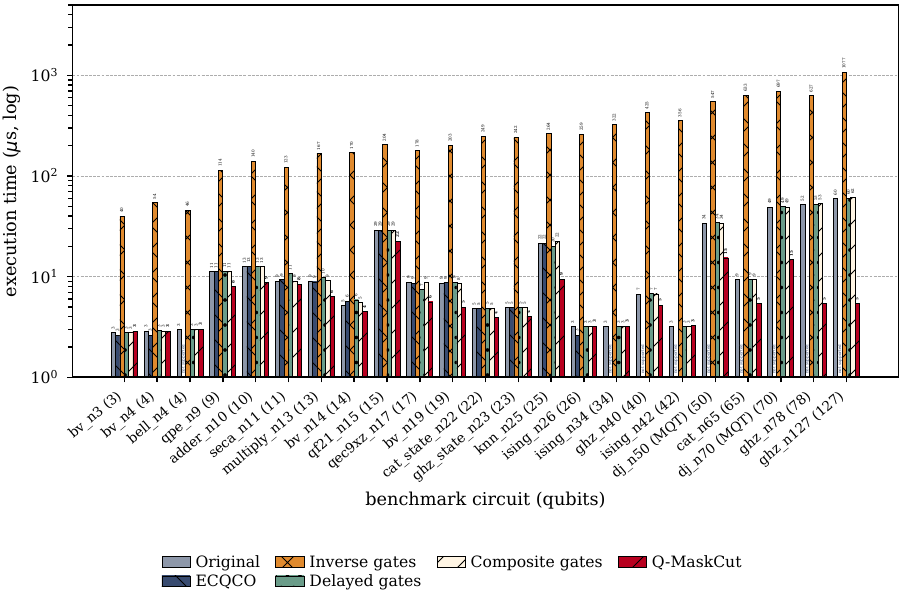}
\caption{Execution time after compilation for FakeMarrakesh (Tier 2: the 23 circuits of the hardware run); Q-MaskCut: the longest-running fragment, as submitted.}
\label{fig:time}
\end{figure}

\begin{figure}[!t]
\centering
\includegraphics[width=\textwidth]{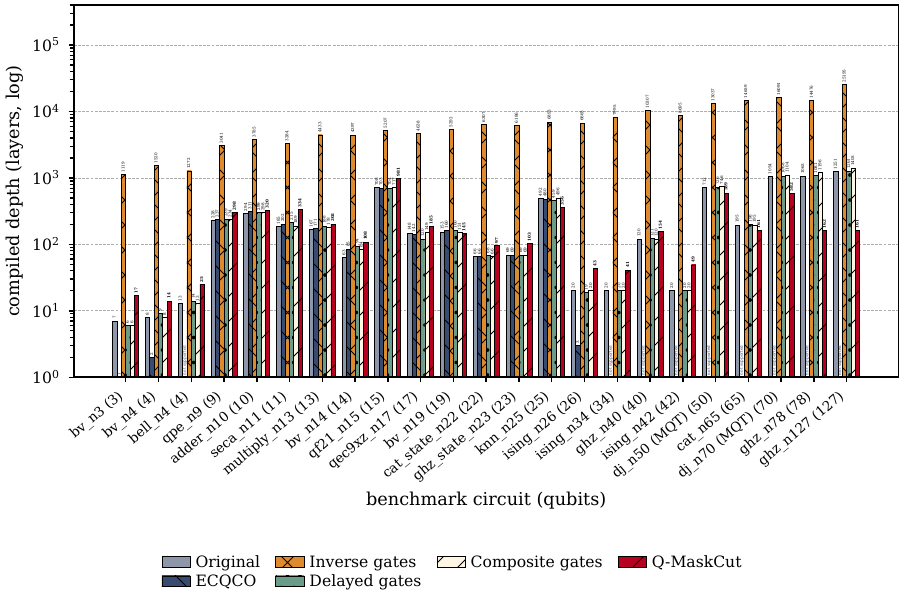}
\caption{Compiled depth (layers) for FakeMarrakesh (Tier 2: the 23 circuits of the hardware run); Q-MaskCut: the longest-running fragment, as submitted, with its masking pulses counted as layers.}
\label{fig:depth}
\end{figure}

Figure~\ref{fig:time} shows the execution time on FakeMarrakesh for Tier 2. The longest-running fragment of Q-MaskCut takes 2.9--22.5~$\mu$s (mean 6.8~$\mu$s), against 2.8--59.9~$\mu$s (mean 15.4~$\mu$s) for the original circuit. It is shorter on 17 of the 23 circuits (for example, 5.4 against 59.9~$\mu$s on ghz\_n127) and longer by at most 0.07~$\mu$s on the other six, all of which are shallow (bv\_n3, bv\_n4, bell\_n4 and the three Ising circuits), because each fragment holds only part of the circuit and the masking fills idle periods. On Tier 1, the means are 7.1 against 9.2~$\mu$s on FakeMarrakesh and 20.8 against 28.2~$\mu$s on FakeMontrealV2; on BV-3, Q-MaskCut is slightly longer than the original (6.58 against 6.13~$\mu$s). Inverse gates take 7 to 112 times as long as the original (mean 310.3~$\mu$s). Composite and delayed gates match the original. Figure~\ref{fig:depth} shows the compiled depth. The masking pulses count as layers, so the deepest fragment of Q-MaskCut is deeper than the original on small circuits (BV-3: 82 against 15 layers), although its mean depth on Tier 2 is lower (223 against 307 layers), for example 161 against 1251 layers on ghz\_n127; inverse gates reach up to 25{,}195 layers.

The main costs of Q-MaskCut are the number of executions and the classical reconstruction. Each Tier-1 circuit is replaced by 7 (one cut) to 91 (three cuts) fragment variants, each run with the full number of shots. The client reconstruction sums $4^K$ terms, so its time grows with the number of cuts and with the number of output outcomes.

\subsection{Hardware run on ibm\_marrakesh}\label{sec:hardware}
The full method of section~\ref{sec:method} was run on the 156-qubit device ibm\_marrakesh for the 23 Tier-2 circuits of 3 to 127 qubits (table~\ref{tab:circuits}), together with the unprotected original of each circuit, with 2048 shots per circuit and per fragment variant (449 circuits in three jobs, 260~s of device time). The original was compiled at optimisation level 3 and submitted without modification. The runtime's own dynamical decoupling and its gate and measurement twirling were disabled for every circuit, so the only added pulses are Q-MaskCut's post-compilation twirl and no-pattern dynamical decoupling (algorithm~\ref{alg:mask}). On the device, fragments of circuits below 40 qubits were limited to 20 wires. Fidelity is the Hellinger fidelity to the exact output; for ising\_n34 and ising\_n42, whose output is too wide and too spread out for an exact reference, it is averaged over 12-qubit blocks against an exact matrix-product-state reference. No device time was used until every protected circuit had passed the noiseless check (fidelity of the decoded output 0.996--1.000). The hardware run was made before the key selection was changed to the random key of algorithm~\ref{alg:protect}: on circuits of up to 14 qubits it used a key chosen among the flip patterns with the largest TVD, and a random key above. The hardware results below are those of that version; the key consists of single-qubit $X$ gates, removed classically, and the hardware run was not repeated with the uniform key.

\begin{table}
\caption{Circuits of the hardware run on ibm\_marrakesh (also the Tier-2 circuits): the Q-MaskCut cut on the device ($K$ cuts; fragment widths include the decoy wires) and the number of fragment variants submitted per circuit.}
\label{tab:circuits}
\centering
\small
\begin{tabular}{lrrlr|lrrlr}
\toprule
circuit & $n$ & $K$ & fragments & variants & circuit & $n$ & $K$ & fragments & variants \\
\midrule
bv\_n3 & 3 & 1 & 2+4 & 7 & ghz\_state\_n23 & 23 & 2 & 9+9+9 & 19 \\
bell\_n4 & 4 & 1 & 3+4 & 7 & knn\_n25 & 25 & 1 & 20+8 & 7 \\
bv\_n4 & 4 & 1 & 4+3 & 7 & ising\_n26 & 26 & 1 & 20+9 & 7 \\
qpe\_n9 & 9 & 2 & 6+7 & 24 & ising\_n34 & 34 & 1 & 20+17 & 7 \\
adder\_n10 & 10 & 2 & 8+6 & 24 & ghz\_n40 & 40 & 1 & 27+16 & 7 \\
seca\_n11 & 11 & 2 & 8+7 & 24 & ising\_n42 & 42 & 1 & 26+19 & 7 \\
multiply\_n13 & 13 & 2 & 13+4 & 25 & dj\_n50 & 50 & 1 & 27+26 & 7 \\
bv\_n14 & 14 & 1 & 9+8 & 7 & cat\_n65 & 65 & 2 & 27+27+15 & 19 \\
qf21\_n15 & 15 & 2 & 8+11 & 24 & dj\_n70 & 70 & 3 & 27+27+21 & 61 \\
qec9xz\_n17 & 17 & 2 & 13+8 & 24 & ghz\_n78 & 78 & 3 & 27+27+27+2 & 31 \\
bv\_n19 & 19 & 2 & 8+7+8 & 19 & ghz\_n127 & 127 & 4 & 4$\times$27+25 & 43 \\
cat\_state\_n22 & 22 & 2 & 8+9+9 & 19 &  & & & & \\
\bottomrule
\end{tabular}
\end{table}

\begin{figure}[!t]
\centering
\includegraphics[width=\textwidth]{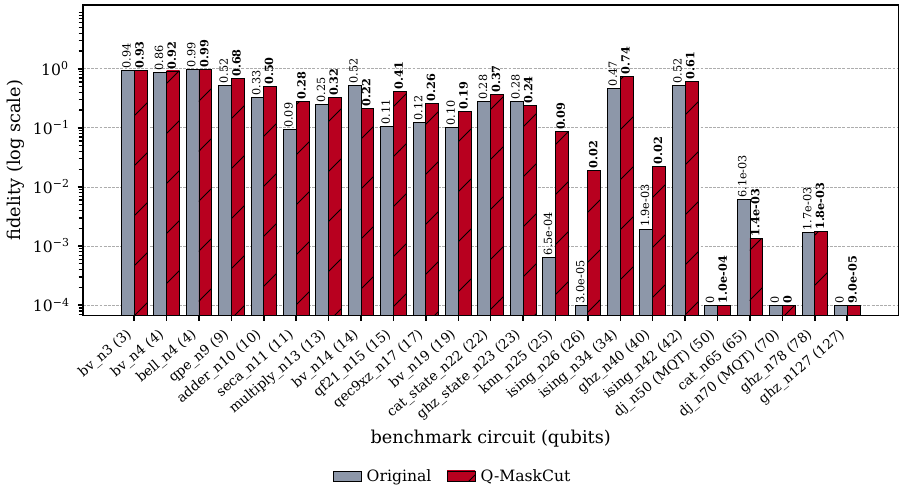}
\caption{Output fidelity on the real IBM quantum computer ibm\_marrakesh, 2048 shots per circuit and per fragment variant. Original: the unprotected circuit, compiled at optimisation level 3 and run as is. Q-MaskCut: the full method; its fragments are run on the device, and the output is then rebuilt and decoded with the key. Logarithmic scale: values below $10^{-4}$, including 0, are drawn at $10^{-4}$ and labelled with their measured value. ising\_n34 and ising\_n42 are too wide for an exact reference and are scored on 12-qubit blocks. The three Ising circuits have a uniform output, so their values reflect the number of shots, not device noise: a noise-free device would score 0.00003, 0.52 and 0.53.}
\label{fig:hw}
\end{figure}

On ibm\_marrakesh (figure~\ref{fig:hw}), each circuit was run once, on one calibration, so the differences below are single measurements. On the 14 circuits of up to 25 qubits, where the device still returns a usable output, Q-MaskCut's decoded output has a higher fidelity than the original by more than 0.02 on 10 (mean 0.457 against 0.386), for example 0.415 against 0.105 on qf21\_n15, 0.280 against 0.093 on seca\_n11, 0.499 against 0.328 on adder\_n10 and 0.683 against 0.524 on qpe\_n9. It is within 0.02 on bv\_n3 and bell\_n4, and lower on two: bv\_n14 (0.216 against 0.524) and ghz\_state\_n23 (0.238 against 0.280). The three Ising circuits do not measure device noise. Their exact output is uniform to within 0.1\% of the maximum entropy on every 12-qubit block, so with 2048 shots even a noise-free device would score 0.00003 on ising\_n26 (whole output), 0.52 on ising\_n34 and 0.53 on ising\_n42 (block-averaged), which is what the original obtained (0.00003, 0.47 and 0.52). Q-MaskCut's higher values there (0.019, 0.744 and 0.611) come from rebuilding the output from several fragment variants of 2048 shots each, not from lower noise, so we do not count them. On the circuits of 40 qubits and more (ghz\_n40, the Deutsch-Jozsa circuits, cat\_n65 and the GHZ circuits of 78 and 127 qubits), both the original and Q-MaskCut returned almost no correct shots (fidelity at most 0.022). Their answers are one or two bitstrings of 40 to 127 bits, every bit of a shot must be correct, and the originals run for 195 to 1251 layers; the device noise therefore exceeds what cutting into fragments of up to 27 wires can recover, although every protected circuit passed the noiseless check.

Table~\ref{tab:overhead} summarises the cost of this run. Each circuit was replaced by 7 to 61 fragment variants (table~\ref{tab:circuits}), each run with 2048 shots, so the 23 protected circuits used 426 variants and 872{,}448 shots, against 47{,}104 shots for the 23 originals (18.5 times more). The longest fragment of Q-MaskCut ran for less time than the original on 19 of the 23 circuits (mean 6.9 against 13.2~$\mu$s).

\begin{table}
\caption{Overhead of Q-MaskCut in the hardware run on ibm\_marrakesh (full method, 23 circuits).}
\label{tab:overhead}
\centering
\small
\begin{tabular}{l>{\raggedright\arraybackslash}p{4.4cm}>{\raggedright\arraybackslash}p{5.6cm}}
\toprule
cost & cause & measured \\
\midrule
circuits per run & each cut multiplies the fragment variants & 7 ($K=1$) to 61 ($K=3$) instead of 1 \\
device shots & every variant runs 2048 shots & 872{,}448 in total against 47{,}104 for the originals \\
classical reconstruction & $4^K$ terms on the client & grows with $4^K$ and with the output size \\
layers on small circuits & masking pulses count as layers & bell\_n4 13 to 31; execution time 3.09 to 3.12~$\mu$s \\
fidelity on some circuits & reconstruction from several fragments & bv\_n14 0.52 to 0.22, ghz\_state\_n23 0.28 to 0.24 \\
\bottomrule
\end{tabular}
\end{table}

\section{Discussion}\label{sec:discussion}

Table~\ref{tab:summary} collects the results of every method under the common protocol.

\begin{table}
\caption{Summary of the comparison (means). normGED: Tier 2, 23 circuits, FakeMarrakesh, against the compiled original, after the user's compilation and after a second compilation by the provider (layout fixed). TVD and fidelity: Tier 1, 11 circuits, 8192 shots; $\mathrm{TVD}_{\mathrm{obf}}$ is the output the provider sees (higher hides more), $\mathrm{TVD}_{\mathrm{deobf}}$ the decoded output (lower is more correct), both without noise; fidelity of the decoded output on FakeMarrakesh. Execution time: Tier 2, for Q-MaskCut the longest-running fragment. Delayed gates and ECQCO were evaluated with their released implementations, which have the problems described in section~\ref{sec:setup}; their rows describe these implementations.}
\label{tab:summary}
\centering
\small
\begin{tabular}{lrrrrrr}
\toprule
 & \multicolumn{2}{c}{normGED} & & & & \\
\cmidrule(lr){2-3}
method & compiled & recompiled & $\mathrm{TVD}_{\mathrm{obf}}$ & $\mathrm{TVD}_{\mathrm{deobf}}$ & fidelity & time ($\mu$s) \\
\midrule
inverse gates & 0.98 & 0.98 & 0.01 & 0.02 & 0.23$^{a}$ & 310.3 \\
composite gates & 0.21 & 0.23 & 0.01 & 0.02 & 0.69 & 15.6 \\
delayed gates & 0.23 & 0.26 & 0.01 & 0.02 & 0.73$^{b}$ & 15.5 \\
ECQCO$^{c}$ & 0.39 & 0.40 & 0.70 & 0.70 & 0.19 & 9.5 \\
Q-MaskCut & 0.88 & 0.42 & 0.94 & 0.02 & 0.78 & 6.8 \\
\bottomrule
\end{tabular}

\smallskip
{\raggedright\footnotesize $^{a}$ The 7 circuits small enough to simulate (original 0.80 on the same circuits). $^{b}$ Fails the operator-equivalence check on all 5 circuits small enough to check (up to 10 qubits; section~\ref{sec:setup}). $^{c}$ Circuits for which its implementation produced a circuit (10 in Tier 1, 14 in Tier 2; original fidelity 0.67 on the same 10); its output could not be decoded with the implementation provided by its authors.\par}
\end{table}

\paragraph{Effect of compilation on inserted gates.} Identity insertions that a compiler can recognise do not survive optimisation level 3, and this removes most of the structural disguise of composite and delayed gates. Masking inserted after compilation, as in Q-MaskCut, escapes the user's compilation but not a second compilation by the provider: recompiling the submitted circuits removed almost all of it and returned normGED from 0.88 to 0.42, whereas the other methods, whose gates are inserted before compilation, changed little. Q-MaskCut's key, decoys and padlocks survive the optimiser, but surviving the optimiser is not the same as hiding the algorithm: part of the remaining normGED comes from the decoy wires, which can be identified from output correlations (see the limitations below). Masking that is to survive recompilation must not be a recognisable identity. Inverse gates survive because the inserted pairs are interleaved with the circuit, but at a depth cost that makes them impractical on current devices.

\paragraph{Structural versus output disguise.} Function-preserving obfuscation, which includes the three ObfusQate schemes, leaves the answer readable by the provider. Hiding the output requires a key that the provider does not hold, together with a way for the user to remove it. Q-MaskCut's bitwise key is removed with a classical XOR, whereas a scheme whose output cannot be decoded, as observed for the implementation of ECQCO provided by its authors, hides the answer from the user as well. Inverse gates keep a high normGED only through added length, which costs fidelity, and composite and delayed gates hide almost nothing after compilation. Among the evaluated methods, Q-MaskCut is the only one that hides the output while the owner recovers it, without lengthening any single execution, at the cost of 7 to 91 executions per circuit; its structural disguise is high only while the provider runs the circuits as submitted, and after recompilation it is comparable to ECQCO's.

\paragraph{Limitations.} Q-MaskCut combines existing techniques and gives no formal security guarantee; normGED and TVD are empirical measures of disguise, and we did not evaluate a provider that analyses the submitted circuits with a dedicated attack. The masking survives neither a second compilation by the provider (section~\ref{sec:recompile}) nor such an analysis. The pulse pairs of each idle gap multiply to the identity and the twirl corrections mirror their partners, so a provider looking for these patterns could recognise and remove them and then try to identify the decoy wires; the decoy Hadamards make each decoy read like a data wire in its marginal statistics, but correlations between output bits can reveal the decoys, and the padlock structure remains in the circuit. Formal security proofs, providers that tamper with the results, and verification of the returned results (for example with trap fragments) are outside the scope of this study. Fragment metadata can also reveal the family of the algorithm~\cite{punch2026post}. In our experiments, all fragments of a circuit ran on the same device, so one provider received every fragment. The benefit that no provider sees the whole circuit holds only when fragments are sent to providers that do not collude, which we did not test. When a single provider receives every fragment, the structural protection rests on the key, the decoys, the padlocks and the masking rather than on the split itself. Q-MaskCut does not detect a provider that alters or replaces fragment results. The cost grows as $4^K$ with the number of cuts, so only circuits that can be cut a few times are practical; our circuits of 40 to 127 qubits needed one to four cuts. The bitwise key hides which bitstring is produced, not how the output is spread, so the TVD it can reach is bounded by the circuit's own output (about 0.71 for bell\_n4); a random key drawn independently of the output reaches less on some circuits (0.36 on bell\_n4), and a single fragment can reveal more than the reconstructed output. With noise, the fidelity of Q-MaskCut is comparable to the original's: it is higher by more than 0.02 on seven of the 11 Tier-1 circuits on FakeMarrakesh and on six on FakeMontrealV2, and within 0.02 on the others (section~\ref{sec:fidelity}). normGED is measured on the whole protected circuit after compilation and before cutting, not on the fragments the provider receives, and it is computed from an upper bound on the GED. Shallow circuits leave little idle time and therefore yield a lower normGED (0.63--0.68 on the Ising circuits of depth 20). The noisy runs of inverse gates on the deepest Tier-1 circuits were too large to simulate. ObfusQate's delayed gates did not pass our operator-equivalence check on any circuit small enough to check (5 Tier-1 circuits of up to 10 qubits) or on 195 of the 200 BV-3 runs, because the gate sequences that its released implementation inserts as identities are not identities (section~\ref{sec:evaluation}); we evaluated the implementation as released. The Pauli twirl and the no-pattern dynamical decoupling were applied to Q-MaskCut only, so we did not separate how much of its fidelity change comes from cutting and how much from these pulses. On ibm\_marrakesh, each circuit was run once, on one calibration. The Ising circuits have a uniform output, so their hardware fidelity reflects the number of shots rather than noise; for ising\_n34 and ising\_n42, it is averaged over 12-qubit blocks, which does not capture correlations between blocks. At 40 qubits and more, neither the original nor Q-MaskCut returned a usable output. Compilation, normGED and depth depend on the device and on its calibration on the day of compilation; the results are for FakeMarrakesh, FakeMontrealV2 and ibm\_marrakesh with fixed seeds. Finally, ECQCO produced circuits for only 14 of the 23 Tier-2 circuits and 10 of the 11 Tier-1 circuits, its own key did not decode its output back to the correct output on seven of these ten, and the comparison is limited to methods with available implementations.

\paragraph{Recommendations.} (i) Measure structural metrics on the compiled circuit, at the provider's highest optimisation level, and again after the provider recompiles the submitted circuit. (ii) Report normGED together with depth and execution time. (iii) Report output disguise together with the correctness of the decoded output. (iv) Check functional correctness on hardware against the exact output, with the unprotected circuit run under the same conditions. (v) Release the protected circuits and raw counts so that results can be reproduced.

\section{Conclusion}\label{sec:conclusion}
This paper compared four existing protection schemes and Q-MaskCut, one example combination of existing techniques, under one protocol in which every protected circuit is compiled before it is measured, with the goal of keeping the provider from learning the user's algorithm and answer. The evaluation covered 11 circuits of 3 to 19 qubits simulated without noise and on two noise models, the compiled structure and execution time of the 23 circuits of 3 to 127 qubits of the hardware run, a Bernstein-Vazirani circuit on 200 random inputs, and a second compilation of the submitted circuits. Compiling first changes the comparison. Composite and delayed gates lose most of their structural disguise, with mean normGED values of only 0.21 and 0.23, and, like inverse gates, they preserve the function, so the provider reads the answer; inverse gates retain a normGED of 0.94 to 1.00 only by making execution 7 to 112 times longer. ECQCO hides the output, but its output could not be decoded with the implementation provided by its authors. Q-MaskCut hides the output (mean TVD without the key 0.94 on Tier 1, 0.66 per fragment on BV-3, with a random key drawn independently of the output), decodes exactly without noise from exact fragment probabilities, and does not lengthen any single execution (mean longest fragment 6.8 against 15.4~$\mu$s). Its compiled normGED of 0.88 comes mostly from the masking added after compilation, and a second compilation removes it: what remains (0.42) comes from the key, decoys and padlocks, which survive the optimiser, although part of it comes from decoy wires that can be identified; it is comparable to ECQCO's, while a second compilation changes the other methods little. Its fidelity with noise is comparable to the original's, and we did not separate the effects of cutting and of the twirl and pulses; its costs are 7 to 91 fragment variants per Tier-1 circuit and a classical reconstruction over $4^K$ terms. These results show that protection methods should be evaluated after compilation and again after recompilation, with structural metrics read together with depth and execution time, and output disguise read together with the correctness of the decoded output.

Q-MaskCut does not introduce a new primitive: it combines a one-time-pad key, decoy wires with padlocks, CutQC wire cutting, and Pauli twirling with no-pattern dynamical decoupling, and its masking reuses pulses that were designed to suppress errors rather than to obfuscate. In future work, we will design a new method that combines full masking with cutting, in which the masking is built specifically for obfuscation. Since a second compilation removes any masking that multiplies to the identity, this masking must change the function of each fragment in a way that only the user can undo, for example additional gates on the decoy wires, random Pauli frames that are not cancelled in the circuit but removed from the output with the key, or masking placed behind mid-circuit measurements, across which the compiler does not optimise; it will be evaluated after a second compilation as well as after the first. Each fragment will be masked under a secret masking key, which selects the dummy operations, qubit relabellings and basis changes inserted into that fragment, so that the masking cannot be recognised by a fixed pattern and the provider sees neither the whole circuit nor the true content of any fragment. We will analyse this method against a provider that actively searches for and removes the masking, test it with fragments sent to different providers, and add a check that lets the user verify the returned fragment results before the reconstructed output is accepted, for example with trap fragments whose outcomes only the user knows.

\funding{This work was supported by Korea Institute for Advancement of Technology (KIAT) grant funded by Korea government (MOTIE) (RS-2026-25607223, Industrial Data-to-Quantum Computing Conversion and Utilization Support Platform).}

\roles{M.I. conceived the original idea, designed the methodology, implemented the algorithms, conducted the experiments, analysed the results and wrote the manuscript. L.H. advised throughout the project, provided guidance on data visualisation, contributed feedback on the experimental results and assisted during manuscript preparation. Y.H. supervised the project, suggested improvements to the methodology and manuscript and contributed to the overall development of the work. All authors reviewed and approved the final manuscript.}

\data{The data and code that support the findings of this study, including the benchmark circuits, the compiled and submitted circuits, the simulated counts, the raw counts returned by ibm\_marrakesh and the analysis scripts, are publicly available at \url{https://github.com/QCL-PKNU/Circuit-Execution-Masking}.}

\section*{ORCID iDs}
\noindent Ing Muyleang \orcid{0009-0008-0551-1754}~\href{https://orcid.org/0009-0008-0551-1754}{0009-0008-0551-1754}\par
\noindent Leanghok Hour \orcid{0009-0001-5670-7481}~\href{https://orcid.org/0009-0001-5670-7481}{0009-0001-5670-7481}\par
\noindent Youngsun Han \orcid{0000-0001-7712-2514}~\href{https://orcid.org/0000-0001-7712-2514}{0000-0001-7712-2514}

\bibliographystyle{iopart-num-doi}   %
\bibliography{refs}

\end{document}